\documentclass[12pt,sort&compress]{article}

\usepackage[dvips]{color}
\usepackage{amsmath}
\usepackage{graphicx} 
\usepackage[dvipsnames]{xcolor}
\usepackage{graphicx}
\usepackage{tabularx}
\usepackage{subfig}
\usepackage{cite}
\usepackage{amsfonts}
\usepackage{tabularx,ragged2e}
\newcolumntype{C}{>{\Centering\arraybackslash}X} 
\usepackage[outercaption]{sidecap}    
\usepackage{hyperref}

\begin{document} 
\title{Non-conserving zero-range processes with extensive rates under resetting}
\date{}
\author{Pascal Grange\\
Department of Mathematical Sciences\\
 Xi'an Jiaotong-Liverpool University\\
111 Ren'ai Rd, 215123 Suzhou, China\\
\normalsize{{\ttfamily{pascal.grange@xjtlu.edu.cn}}}}
\maketitle

\begin{abstract}
We consider a non-conserving zero-range process with hopping rate proportional to the number of particles at each site. Particles are added to the system with a site-dependent creation rate, and vanish with a uniform annihilation rate. On a fully-connected lattice with a large number of sites,  the mean-field geometry  leads to a negative binomial law for the  number of particles at each site, with parameters depending on the hopping, creation and annihilation rates. This model can be mapped to population dynamics (if the creation rates are reproductive fitnesses in a haploid population, and the hopping rate is the mutation rate).  It can also be mapped to a Bianconi--Barab\'asi model of a growing network with random rewiring of links (if creation rates are the rates of acquisition of links by nodes, and the hopping rate is the rewiring rate). The steady state has recently been worked out and gives rise to  occupation numbers that reproduce Kingman's house-of-cards model of selection and mutation. In this paper we solve the master equation using a functional method, which yields  integral equations satisfied by the occupation numbers. The occupation numbers are shown to forget initial conditions at an exponential rate that decreases linearly with the fitness level. Moreover, they can be computed exactly in the Laplace domain, which allows to obtain the steady state of the system under resetting. The result modifies the house-of-cards result by simply adding a skewed version of the initial conditions, and by adding the resetting rate to the hopping rate.  
\end{abstract}

 \pagebreak

\tableofcontents

\section{Introduction}

 Non-conserving systems of particles on a fully connected lattice are useful 
  models to describe out-of-equilibrium systems with varying 
 levels of fitness.
 In the context of population dynamics, particles represent individuals in a haploid population, 
 and fitness is understood as reproductive fitness.
 In the context of network science, particles at a node represent links to this node (this 
mapping from configurations of links to configurations of particles was
  used in the construction of the Bianconi--Barab\'asi model of a growing network \cite{BB}), and fitness is
  understood as the ability to acquire new links.\\

The accumulation of particles at sites of higher fitness results from a  selection 
 process. This effect can be attenuated by other processes: vanishing of particles, 
 and production of particles of random fitness. This production process corresponds to 
 mutation in a population (and it corresponds to random
 rewiring of links in a network). One may ask whether a finite proportion of the particles can condense at the highest 
 fitness level, provided death and mutation rates are low enough. In \cite{ZRPSS}, such a model of particles on a lattice was mapped
 to a non-conserving zero-range-process (ZRP) \cite{spitzer5interaction,jamming,EvansBraz,DrouffeSimple} with extensive 
  hopping rates, describing the occupation 
 number of each fitness level.
  The zero-range process is well-known to give rise to condensation, whose
 emergence and dynamics are well studied in the case of 
 decreasing hopping rates 
 \cite{bialas1997,majumdarMass,ZRPReview,godreche2003,grosskinsky2003condensation,condensationInhomogeneous,coarsening,mailler2016condensation}.
  {{Non-conserving ZRPs coupled to reservoirs through open boundaries are studied in \cite{levine2005zero}, and stationary currents 
 can be induced by asymmetry of the hopping rate at a defect \cite{cirillo2017stationary}.}}\\

 The steady state of the non-conserving ZRP
 was worked out  in \cite{ZRPSS} from a master equation by adapting techniques from \cite{angel2005critical,nonConserving}.
  The expectation value of the {{occupation number of each fitness level in the steady state}} is the sum of  two terms: a skewed version of the  
  mutant density, and  a term that approaches an atom supported at maximum fitness
  in a certain regime of  low   death and mutation      rates.  These two terms evoke the
 result of the deterministic measure-valued house-of-cards model introduced by Kingman in population dynamics to 
 describe the balance of selection and mutation \cite{KingmanSimple} (which assumes non-overlapping generations
  and large populations).   As the steady state of the system is unique, the  occupation numbers
 forget the initial conditions. However, the rate at which they do so may depend on the fitness.\\ 

  On the other hand, a system may be put in contact with initial conditions at random times by undergoing stochastic resetting.
    This process resets the system to its initial configuration or to an ensemble of resetting configurations \cite{evans2011diffusion,evans2011optimal},
  and keeps it from its stationary state. The system under resetting can feature a non-equilibrium stationary 
  state, depending on the parameters of the resetting process \cite{evans2014diffusion}. {{Characterisation of  such a non-equilibrium stationary sate has
  recently become a major focus of activity in statistical physics \cite{majumdar2015dynamical,maes2017induced,masoliver2019anomalous}. The field enjoys a  broad range of applications including RNA polymerisation processes \cite{roldan2016stochastic,lisica2016mechanisms}, active matter \cite{scacchi2018mean,evans2018run,masoliver2019telegraphic} and randomised searching problems \cite{kusmierz2014first} (see the review \cite{stochasticReview} and references therein).}}\\

  Moreover, the probability 
 law of configurations of a system under resetting is related to the probability 
 law of the ordinary system (without resetting) 
 by an integral equation, obtained from a renewal argument \cite{evans2011diffusion,evans2018run,lapeyre2019stochastic,gupta2019stochastic,basu2019long,basu2019symmetric,grange2019entropy}.
 In particular, the probability law at steady state of the system under resetting 
 can be worked out from the Laplace transform of the 
 probability law at  steady state of the ordinary system.
 In the  present model of particles,  a resetting event corresponds to a massive
 extinction of population or to a destruction of links in the network, restoring the 
  system to its initial configuration.\\

 
 {{In this paper we  characterise the behaviour
 of a non-conserving ZRP at finite time (in particular, we work out how fast the occupation numbers 
 forget the  initial conditions), in order to characterise the steady state of the system under resetting.}}  
   In Section 2 we review the model and introduce notations, recalling the interpretations
 of the parameters in population dynamics and network science. In Section 3 we 
  introduce the  generating function of the model and work out the PDE it satisfies,
 based on the master equation. 
  This PDE is non-linear, because of the presence of the average density of the system
 as a coefficient. This situation occurs in balls-in-boxes models, such as the backgammon model\cite{ritort1995glassiness,franz1995dynamical,godreche1996long,bialas1997} (in which
  a fixed number of balls are distributed among boxes, and 
 the energy of a configuration is given by the number of empty boxes). 
 In Section 4 we adapt the functional approach taken in \cite{franz1996glassy} to 
 solve the master equation: the moment-generating function is considered as a functional of the density,
 which allows to solve the PDE after a change of variables.
  A consistency condition is imposed, which yields an integral equation satisfied by the density.
 This equation is solved in the Laplace domain. In Section 5, this solution is used 
 to  work out the expectation values of the occupation numbers at each fitness level 
 of the system undergoing stochastic resetting to its initial configuration at a fixed rate.\\

\section{Review of the non-conserving ZRP with extensive rates}\label{model}


\subsection{Description of the system and quantities of interest}

{{
  Consider a fully connected lattice with a large number of nodes (or states). Each of these states can be occupied 
  by a certain number of particles. 
  Moreover, each  state carries a certain value, called its fitness level. This value is interpreted  either in the sense 
 of  reproductive fitness (cell division rate) in a haploid population, or in the sense of the rate of acquisition of new links by the node in a network (see Table \ref{laTable}
 for a summary of the parameters with their interpretations).}}\\

{{The occupation number of each state is a random quantity, evolving through the following processes:\\
1. hopping to another uniformly drawn state (the particles are independent 
 random walkers on the fully connected set of states in the system),\\
2. annihilation (at a uniform rate),\\
3. creation (at a rate that increases with the fitness level).\\ 
 The above three processes were already considered in \cite{ZRPSS}. In this paper we will moreover consider stochastic resetting 
 of the system to an empty configuration: massive extinctions occur at random times. When a massive extinction happens, 
 all particles are  instantaneously annihilated (the set of states stays fixed).}}\\

{{
The main quantities of interest are the occupation numbers of each fitness level (the sums of the occupation numbers of the
 states at this fitness level).
 The fitness is assumed to be bounded, with a maximum of 1, as in the house-of-cards model  \cite{KingmanSimple}.
 We are going to review the formulation of the model with a discrete set of $L$ regularly-spaced fitness levels  in the interval $]0,1]$.
 In the large-$L$ limit, the hopping process will be described in terms of the average density of the system using a mean-field 
 approach, and the fitness levels will approach a continuum. The evolution of the probability 
  law of the occupation number of a given fitness level will result in the master equation (Eq. \ref{evolPDE})
 derived in \cite{ZRPSS}, which is valid between massive extinctions.}}


\subsection{Master equation}

{{
 Let us  consider a large number $L$ of regularly spaced fitness levels  in the interval $]0,1]$. Let us denote by $v_k$ the number of states
 at  fitness level $k/L$  (for each $k$ in $[1..L]$), with 
 \begin{equation}\label{vl}
v_k = \mathrm{max}\left( \left[ \frac{1}{L}q\left(\frac{k}{L}\right) V \right], 1\right) = \frac{1}{L}q\left(\frac{k}{L}\right) V + \zeta_k,
 \end{equation}
 where $V$  denotes a large integer, satisfying $V>>L$.
  The symbol $q$ denotes a fixed probability density, defined on the  continuum $[0,1]$, one of the parameters of the model (which corresponds 
 to the mutant density in Kingman's house-of-card model of population dynamics \cite{KingmanSimple}). 
  Let us assume that the density $q$ vanishes at maximum fitness:
\begin{equation}\label{qCond}
 q(1) = 0.
\end{equation}
so that there is only one state at the maximum fitness level.  Square brackets denote the integer part, and $\zeta_k$ is a number in $[-1,1]$.}}\\

 {{
 The total number of states in the system is  of order $V$ (it is not exactly equal to $V$ due to the integer-part prescription used in Eq. \ref{vl} to obtain integer numbers 
 from the density $q$): 
 \begin{equation}\label{totalNumberofStates}
\sum_{k=1}^L  v_k = \left( \frac{1}{L} \sum_{k=1}^L q\left(\frac{k}{L}\right)  + \frac{1}{V}\sum_{k=1}^L \zeta_k \right)V.
 \end{equation}
 As  $|\zeta_k|<1$, the absolute value of the second sum is lower than $L/V$.}}\\

 {{ 
  For each integer $k$ in $[1..L]$, let us label the $v_k$ states with fitness $k/L$ by integers in $[1..v_k]$, and denote by $\nu(k,s)$ the 
 occupation number of the state labelled by $s$. The total number $n_k$ of particles at fitness level $k/L$ is expressed as
\begin{equation}\label{nkexpr}
n_k = \sum_{s=1}^{v_k} \nu(k,s).
\end{equation}
 Let us describe the hopping process, to review how it induces a ZRP with extensive rates between fitness levels.
  The particles are independent random walkers on the fully connected set  of states described by 
 Eq. \ref{vl}. Each of these walkers hops 
 at a fixed rate $\beta$ to a uniformly  drawn target state. Consider the hopping processes that increase the occupation number of  the set of states
 at a fixed level of fitness $m/L$, for some integer $m$ in $[1..L]$.
Let us  express the total rate of these processes, using Eqs \ref{vl},\ref{totalNumberofStates} and \ref{nkexpr}:
\begin{equation}\label{discHoppingRate}
\begin{split}
 \beta \sum_{k\neq m} \sum_{s=1}^{v_k}\nu(k,s) \times \frac{v_m}{\sum_{j=1}^L v_j} &= \beta \left( \sum_{k\neq m} n_k\right) 
\times \frac{\frac{1}{L}q\left(\frac{m}{L}\right)  + \frac{\zeta_m}{V}}{\frac{1}{L} \sum_{j=1}^L q\left(\frac{j}{L}\right)  + \frac{1}{V}\sum_{j=1}^L \zeta_j  }\\
&= \beta \left( \left(\sum_{k=1}^L n_k \right)- n_m \right)  \times \frac{\frac{1}{L}q\left(\frac{m}{L}\right)  + \frac{\zeta_m}{V}}{\frac{1}{L} \sum_{j=1}^L q\left(\frac{j}{L}\right)  + \frac{1}{V}\sum_{j=1}^L \zeta_j }\\
&=  \beta \left(  \frac{1}{L}\sum_{k=1}^L n_k  - \frac{ n_m}{L} \right)  \times \frac{ q\left(\frac{m}{L}\right)  + \zeta_m\frac{L}{V}}{\frac{1}{L} \sum_{j=1}^L q\left(\frac{k}{L}\right)  + \frac{1}{V}\sum_{j=1}^L \zeta_j  }\\
&\xrightarrow[V\to \infty]{}   \beta \left(  \frac{1}{L}\sum_{k=1}^L n_k - \frac{ n_m}{L} \right)  \times \frac{ q\left(\frac{m}{L}\right) }{\frac{1}{L} \sum_{j=1}^L q\left(\frac{k}{L}\right)   }.\\
\end{split}
\end{equation}
 }}

{{ 
 At fixed $L$, for a given configuration of occupation numbers, we can define a function $n^{(L)}$ on the interval $]0,1]$ by   
\begin{equation}
 l \in ]0,1] \mapsto n^{(L)}(l) = n_{[lL]},
\end{equation}
which is piecewise constant.  In the large-$L$ limit, let us denote the limit of  $n^{(L)}(l)$  by $n(l)$. Repeating this procedure for all configurations of occupation numbers  induces
 a continuous one-parameter family of random variables, the random occupation numbers of fitness levels. Let us denote by $p_l(.,t)$ the probability law of $n(l)$ at time t: 
\begin{equation}\label{pl}
\forall l \in ]0,1],\;\;\;\;\;p_l(n,t):= P(n(l)= n \;\mathrm{at}\;\mathrm{time}\;t).
\end{equation}
The average density defined of the system can therefore be expressed in the large-$L$ limit as an integral:
\begin{equation}\label{densities}
\overline{n_l}(t):= \sum_{n\geq 0 } n p_l( n,t),\;\;\;\;\;\rho(t) := \int_0^1 \overline{n_l}(t) dl.
\end{equation} 
}}
 
 {{ Moreover, 
  the denominator in Eq. \ref{discHoppingRate} is a Riemann sum which goes to 1 when $L$ goes to infinity, because $q$ is a probability density. 
 In the continuum limit, we consider a fixed value of the fitness $l$ in $]0,1]$, and in the rate of hopping processes to states with fitness $m/L$, we take the limit 
 in which both integers  $L$ and $m$ go to infinity, while the ratio $m/L$ goes
 to $l$.  Moreover, in the large-$L$ limit, each fitness level is connected by the hopping process to 
 a large number of other fitness levels. In this limit, we are therefore in a situation that is suitable for a mean-field approximation, as  in the dynamics of  urn models 
  \cite{godreche2003,urn1,urn2}.
  We  therefore approximate the factor $(\sum_{k=1}^L n_k)/L -n_m/L$ by the average density $\rho$. Neglecting  the term $n_m/L$  (which depends on the destination level) corresponds to the fact that hopping processes relate states with distinct fitness levels with a probability that goes to one when $L$ goes to infinity.}}\\

{{
Similarly, let us express the total rate of the hopping processes that decrease the occupation number of the the set of states
  of fitness $m/L$:
\begin{equation}\label{discHoppingRateDecrease}
\begin{split}
 \beta  \sum_{k\neq m} \sum_{s=1}^{v_m}\nu(m,s) \times \frac{v_k}{\sum_{j=1}^V v_j}&= \beta n_m  \frac{\sum_{k\neq m} v_k}{\sum_{j=1}^L v_j}\\
&=   \beta n_m \left(  1 -  \frac{v_m}{\sum_{j=1}^L v_j}  \right)\\
&=  \beta n_m \left(  1 -  \frac{\frac{1}{L}q\left(\frac{m}{L}\right)  + \frac{\zeta_m}{V}}{\frac{1}{L} \sum_{j=1}^L q\left(\frac{j}{L}\right)  + \frac{1}{V}\sum_{j=1}^L \zeta_j  }  \right)\\
& \xrightarrow[V\to \infty]{} \beta n_m \left(  1 -  \frac{\frac{1}{L}q\left(\frac{m}{L}\right) }{\frac{1}{L} \sum_{j=1}^L q\left(\frac{j}{L}\right)}   \right).
\end{split}
\end{equation}
 The factor of $n_m$ is the only dependence of this rate on the occupation numbers, which is the defining property of the ZRP. Taking again the limit 
 of large $L$ and $m$, with $m/L=l$ fixed, the last factor in the above expression goes to 1. The total rate of the hopping process from 
 the states at level $l$ therefore goes to $\beta n(l)$.}}\\

{{
We can therefore combine the continuum limit of Eqs \ref{discHoppingRate} and \ref{discHoppingRateDecrease}, to express the contribution of the hopping process to the master equation for the occupation number at fitness level $l$:
 \begin{equation}
\begin{split}
 \forall n \in \mathbf{N},\;\;\;\left(\frac{dp_l(n,t)}{dt} \right)_{\mathrm{hopping}} = &\beta \rho(t) q(l) p_l(n-1,t) \theta(n) - \beta  \rho(t) q(l) p_l(n,t) \\
   &   - \beta n p_l(n,t)  \theta(n) + \beta (n+1) p_l(n+1,t),\\
\end{split}
 \end{equation}
where the first two terms account for hopping processes to level $l$ (one increasing the occupation number from $n-1$ to $n$, with the step-function 
 factor $\theta(n)=\mathbf{1}(n>0)$ ensuring that this term exists only for $n>0$, the other increasing the occupation number from $n$ to $n+1$). The last 
 two terms account for hopping  processes from level $l$ (the third term decreases the occupation number from $n$ to $n-1$, again with a
 factor $\theta(n)$ even though it is redundant due to the factor of $n$, and the fourth term decreases the  occupation number from $n+1$ to $n$).
 The assumption of Eq. \ref{qCond}  implies that the occupation level of the maximum fitness level is never increased by the hopping process. Moreover, the 
 factor of $q(l)$ in the first two  terms  yields the interpretation of the density $q$ in population dynamics as the mutant density \cite{KingmanSimple,ZRPSS}: when a new mutant
 is added to the population, its fitness is drawn from the probability distribution $q$. An accumulation of a large fraction 
 of the total population at maximum fitness could therefore only be attributed to selection, and not to beneficial mutations). The prescription of Eq. \ref{vl} ensures that the 
 density of states goes to the fixed density $q$ in the limit of a large number of nodes and a large number of fitness levels (it is not the only prescription with such a limit: there needs to be a non-zero number of states of maximum fitness to support particles, but any fixed number independent of $L$ and $V$ would do). }}\\

{{
Let us describe the annihilation and creation processes directly in the continuum limit, as they
 do not couple occupation numbers of distinct fitness levels. All particles in the system 
 are assumed to be annihilated at the same rate, denoted by $\delta$. The total annihilation rate at a fixed fitness level  $l$
 is therefore proportional to the occupation number of this level. 
 The annihilation process  induces the following two terms in the master equation:
 \begin{equation}
 \forall n \in \mathbf{N},\;\;\;\left(\frac{dp_l(n,t)}{dt} \right)_{\mathrm{annihilation}} =  - \delta n p_l(n,t) \theta(n)  + \delta (n+1) p_l(n+1,t),
 \end{equation}
where the first term corresponds to decreasing the occupation number from $n$ to $n-1$, with the constraint $n>0$,
 and the second term corresponds to decreasing the occupation number from $n+1$ to $n$.}}\\

 {{
 The rates of the creation process are biased by the fitness level: we consider the rate of creation
 of particles at level $l$ to be $l(n(l) + 1)$. The factor of $l$ models the selection process: rates of creation increase with 
 fitness. The  other factor is the occupation number of the fitness level {\emph{after}} the creation of a new particle. This amounts
 to modelling the creation process as an annihilation process reversed in time. In the population-dynamics interpretation this
  implies that there is a spontaneous generation
 of particles at empty fitness levels (an operator adds an individual at fitness level $l$ if this level becomes empty, to keep the selection process going).
  In the network interpretation of the model, this is quite natural as nodes can acquire 
 links from other nodes even if they do not have any yet.
 The creation process therefore induces the following two terms in the master equation:
\begin{equation}
 \forall n \in \mathbf{N},\;\;\;\left(\frac{dp_l(n,t)}{dt} \right)_{\mathrm{creation}} =  l  n  p_l(n-1,t) \theta(n)  -  l(n+1)  p_l(n,t),
 \end{equation}
 where the first term  corresponds to increasing  the occupation number  from $n-1$ to $n$, with the constraint $n>0$, and the
 second term corresponds to increasing the occupation number from $n$ to $n+1$. Setting the maximum level of fitness to $1$ is equivalent 
 to setting the time scale of the process: starting from a configuration with no particles, the rate of creation of particles
 at the highest fitness level equals 1.}}\\

\begin{table}[!ht]
\setlength\extrarowheight{2pt} 
\begin{tabularx}{\textwidth}{|p{1.5cm}|p{2.3cm}|C|C|C|}
   \hline
   Symbol & Values &Particles & Network & Population \\
   \hline
   $l$ &   $l\in ]0,1]$ & rate of {\hbox{production}}  of particles       & rate of {\hbox{acquisition}} of links {{(fitness of a node in a network)}} & rate of cell division in a haploid population  {{(reproductive fitness)}}\\
 \hline
   $q$ &   probability density on $[0,1]$, satisfying $q(1)=0$ & density of states     &  density of states     &   mutant fitness            \\
\hline
  {{$n(l)$}} &  random integer & total  {\hbox{occupation}} number in  states  at level $l$ & total number of links to nodes at level $l$ & total population of fitness $l$\\
   \hline
 $\delta$ &  $\delta > 0$ & vanishing rate      & rate of {\hbox{disappearance}} of links             & death rate     \\
\hline
   $\beta$ &  $\beta >0$ & hopping rate &  rewiring rate            & mutation rate         \\
 \hline
\end{tabularx}
\caption{Table of notations and interpretations for the parameters of the model, {{in the limit of a continuum of fitness levels}}.}
\label{laTable}
\end{table}

%


 {{If we disregard the resetting of the occupation numbers to $0$ induced by massive extinctions, or alternatively 
 consider the evolution of the system between resetting times, the probability law of the occupation number 
 of  fitness level $l$
   satisfies the following master equation \cite{ZRPSS}, induced by the hopping, annihilation and creation processes (paramatrised by the quantities listed in Table \ref{laTable})}}:
\begin{equation}\label{evolPDE}
 \begin{split}
 \frac{dp_l(n,t)}{dt} = \;& \theta( n ) \left\{  \left( \beta \rho(t) q\left( l \right) +  l  n \right)p_l(n-1,t) 
  - ( \delta  +  \beta ) n p_l(n,t)    \right\}\\
   &+   (  \beta+ \delta)(n+1)p_l(n+1,t)  -  \left( \beta \rho(t) q\left(  l \right) +   l ( n+1)\right)p_l( n,t), \;\;\;\forall n \geq 0.
 \end{split}
 \end{equation} 


 At steady state ({{in the absence of resetting events}})  the probability distribution of particles at each fitness level is negative
 binomial. This was established in \cite{ZRPSS} by putting the l.h.s of Eq. \ref{evolPDE} to zero. 
  An expression for the steady-state 
 occupation numbers $\overline{n(l)}(\infty)$ was derived. It consists of two terms:
 a skewed version of the measure $q$, and a term that gives gise to an atom at maximum fitness
 when the sum of vanishing and hopping rates $\delta + \beta$  decreases to 1:
\begin{equation}\label{steadyOrdinary}
  \overline{n}_{l}(\infty) =  \frac{l}{\beta + \delta -l} + 
     \frac{ \beta\rho(\infty)q(l)}{\beta + \delta - l},
 \end{equation}
\begin{equation}
  \rho(\infty)  =    \int_0^1 \frac{l}{\delta + \beta -l} dl\left( 1 - \beta \int_0^1 \frac{q(l) dl}{\delta +\beta -l}\right)^{-1},
\end{equation}
\begin{equation}\label{critical}
   {\mathrm{provided}}\;\;\; \;\delta + \beta > 1 \;\;\;\;{\mathrm{and}}\;\;\;\;\;
\beta < \beta_c = \left(  \int_0^1 \frac{q(l) dl}{\delta +\beta -l}\ \right)^{-1}. 
 \end{equation}
The structure of this expression, together with the critical value of the hopping rate, is  
 very reminiscent of the steady state of the house-of-cards model \cite{KingmanSimple}. The steady state is unique 
 and does not depend on the initial conditions  $(\overline{n}_{l}(0))_{l\in]0,1]}$. In the rest of the paper we will assume
 that the rates $\delta$ and $\beta$ satisfy the conditions of Eq. \ref{critical} leading to a steady state.
   We are going to study the master equation at finite time in order to work out how fast 
 each fitness level forgets initial conditions, {{in the absence of resetting events}}.
   {{The solution will be used to address resetting events in Section \ref{resetting}.}}





\section{From the master equation to the generating function}

   Let us introduce the generating function, which  characterises  the probability law of all the {\hbox{configurations}} of occupation numbers:
\begin{equation}
 H(J,l,t) :=  \sum_{n\geq 0} p_l(n,t) J^n.
\end{equation} 
  In particular, the average occupation numbers and the density parameter can be expressed at time $t$ in terms of $H$ as follows:
 \begin{equation}\label{densities}
 \overline{n_l}(t) = \frac{\partial H}{\partial J}|_{J=1}(J,l,t), \;\;\;\;\;\;\;\;\rho(t) = \int_0^1 \left( \frac{\partial H}{\partial J}|_{J=1}(J,l,t) \right) dl.
 \end{equation}

  The  master equation induces a  PDE in variables $t$ and $J$, satisfied by the generating function at each level $l$. 
   Let us write it  with the terms  in  the same order as in Eq. \ref{evolPDE}:
\begin{equation}
 \begin{split}
 \frac{\partial H}{\partial t} =& \beta \rho q( l ) JH  + lJ^2\frac{\partial H}{ \partial J} + lJ H - (\beta+\delta) J\frac{\partial H}{\partial J}\\
  &+ (\beta+\delta)  \frac{\partial H}{ \partial J} -\beta \rho q(l) H  - l J \frac{\partial H}{ \partial J} - lH.
 \end{split}
\end{equation} 
 Because of the dependence of the coefficients on the density $\rho$ {{(defined in terms of the probability laws of occupation numbers in Eq. \ref{densities})}},
 the PDE is non-linear: 
\begin{equation}\label{timeGener}
 \frac{\partial H}{\partial t} = ( 1-J) \left( ( -l -\beta \rho q(l) )H+ (1+\zeta -lJ)\frac{\partial H}{\partial J}  \right),
\end{equation}
where 
\begin{equation}\label{zetaDef}
\beta + \delta = 1+\zeta,
\end{equation}
 with $\zeta > 0$ (this assumption, Eq. \ref{critical}  ensures that the sum of the vanishing and hopping rates is high enough 
 to prevent an exponential divergence of the occupation number at maximum fitness). 
  The non-linearity can be addressed using the functional method 
   (see  \cite{franz1996glassy} for an application of this method to the backgammon model). 
  We are going to consider the generating function  as a functional of the density $\rho(t)$, and to impose a 
  closure condition through the general definition (Eq. \ref{densities}) of the density.

  \section{The generating function as a functional of the density}
 \subsection{Change of variables}
 To work out the functional dependence of the generating function on the density, 
  let us look for a change of variables  that would transform Eq. \ref{timeGener} into an ordinary differential equation
  in the time variable. The equation contains no derivative w.r.t. the fitness variable $l$, so we
  denote the new variables by $(v,l,t)$ and look for a function $J(v,l,t)$, so that 
 the generating function can be expressed in the new variables as:
\begin{equation}\label{functionalDef}
 \hat{H}( v, l, t ) := H( J(v,l,t), l, t ).
\end{equation}
 Taking the derivative of Eq. \ref{functionalDef}  w.r.t. time we obtain:
 \begin{equation}
 \frac{\partial \hat{H}( v, l, t )}{\partial t} = \frac{\partial H( J(v,l,t), l, t )}{\partial J}  \frac{\partial J(v,l,t)}{\partial t } + \frac{\partial H( J(v,t), l, t )}{\partial t }.
 \end{equation}
 Inspecting Eq. \ref{timeGener}, we  impose the following condition, in order to match the derivatives
 of the moment-generating function w.r.t. $J$:
 \begin{equation}\label{imposed}
\frac{\partial J}{\partial t } = (J-1)(1+\zeta -lJ).
 \end{equation}
 As the parameter $\rho$ is a function of time only, and the density $q$ is a function of $l$ only,
   Eq. \ref{timeGener} can be rewritten in the variables $(v,l,t)$ as
  \begin{equation}\label{ODEtrans}
 \frac{\partial \hat{H}( v, l, t )}{\partial t} = \left(1-J(v,l,t) \right)\left(-l-\beta \rho(t) q(l)\right) \hat{H}( v, l, t ).
 \end{equation}
 We have to integrate  Eq. \ref{imposed} w.r.t. time to work out the change of variables.
 The inverse of the  r.h.s. is a 
  rational function of the parameter $J$, with two poles, at $J=1$ and $J =( 1+\zeta)/l>1$, for any fixed $l$ in ]0,1[. 
 Decomposing it into simple elements yields:
 \begin{equation}
 \frac{1}{ (J-1)(1+\zeta -lJ)} = \frac{1}{1+\zeta-l}\left( \frac{1}{J-1}-\frac{l}{lJ - 1 -\zeta} \right).
 \end{equation}
 For our purposes $J$ is in the interval $[0,1]$. The quantity $lJ - 1 -\zeta$ is therefore negative,
 and we can  rewrite Eq. \ref{imposed} as
 \begin{equation}
   \frac{1}{1+\zeta -l}\left( \frac{\partial \log( 1 - J )}{\partial t}  -  \frac{\partial \log( 1 +\zeta -lJ )}{\partial t}  \right)= 1.
 \end{equation}
 Let us integrate this relation w.r.t. time and denote by $v$  the integration constant
  \begin{equation}
    \log\left( \frac{1 +\zeta -lJ(v,l,t) }{ 1 - J(v,l,t)}  \right)= v + ( l-1-\zeta) t.
 \end{equation}
 The change of variables from $(J,l,t)$ to $(v,l,t)$ is therefore 
 defined by the relation:
   \begin{equation}
    \frac{1 +\zeta -lJ(v,l,t) }{ 1 - J(v,l,t)}   =  e^v e ^{ ( l-1-\zeta) t}.
 \end{equation}
 We can   express the factor of $(1-J)$ needed in Eq. \ref{ODEtrans} in the 
 variables $(v,l,t)$ as:
  \begin{equation}\label{explicitChange}
    1 - J(v,l,t)   =  \frac{l-1-\zeta}{l- e^v e^{ ( l-1-\zeta) t}},
 \end{equation}
  which finally yields an explicit form in terms of the variables $(v,l,t)$ for the time-evolution of the generating function:
 \begin{equation}
 \frac{1}{\hat{H}( v, l, t )}\frac{\partial \hat{H}( v, l, t )}{\partial t} =  \frac{l-1-\zeta}{l- e^v e^{ ( l-1-\zeta) t}}
\left(-l-\beta \rho(t) q(l)\right).
 \end{equation}

 Integrating w.r.t. time yields an expression of the generating function as a functional of the density:
 \begin{equation}\label{solInv}
 \log\left( \frac{ \hat{H}( v, l, t )}{ \hat{H}( v, l, 0 )} \right)=(l-1-\zeta) \int_0^t \frac{-l-\beta \rho(s) q(l)}{l- e^v e^{ ( l-1-\zeta) s}} ds.
 \end{equation}

 We have to transform back to the variables $(J,l,t)$.
 Let us rewrite the  functional relation (Eq. \ref{functionalDef}), and
 the explicit change of variables Eq. (\ref{explicitChange}), at time $0$:
 \begin{equation}
  \hat{H}( v, l, 0 ) := H( J(v,l,0), l, 0 ),
 \end{equation}
 \begin{equation}
    1 - J(v,l,0)   =  \frac{l-1-\zeta}{l- e^v}.
 \end{equation}
 From the change of variables at time $t$ (Eq. \ref{explicitChange}) we have
 an expression of $e^v$ in terms of the quantities $J(v,l,t)$, $l$ and $t$,
 from which we find
\begin{equation}
   J(v,l,0)   = 1- \frac{l-1-\zeta}{l-\left( l + \frac{1+\zeta - l}{1-J(v,l,t)} \right) e^{(1+\zeta - l )t} }.
\end{equation}
Going back to Eq. \ref{solInv}, we obtain the desired equation in the variables $(J,l,t)$:
\begin{equation}\label{logH}
 \begin{split}
 \log&\left(  H( J, l, t )\left(  H\left( 1- \frac{(1-J)(l-1-\zeta)}{ l(1-J)-\left( l(1-J) + (1+\zeta - l) \right) e^{(1+\zeta - l )t}}, l, 0 \right) \right)^{-1}
\right)\\
&=(l-1-\zeta) \int_0^t \frac{-l-\beta \rho(s) q(l)}{l-  \frac{1+\zeta - l J }{1-J}e^{ ( l-1-\zeta) (s-t)}} ds\\
&=  (l-1-\zeta)(1-J) \int_0^t \frac{-l-\beta \rho(s) q(l)}{l(1-J)-  (1+\zeta - l J )e^{ ( l-1-\zeta) (s-t)}} ds \\
\end{split}
\end{equation}
hence
 \begin{equation}
\begin{split}
 H( J, l, t ) =  & H\left( 1- \frac{(1-J)(l-1-\zeta)}{ l(1-J)-\left( l(1-J) + (1+\zeta - l) \right) e^{(1+\zeta - l )t}}, l, 0 \right)\\
& \;\;\;\;\;\times \exp\left(   (l-1-\zeta)( 1 - J)  C[\rho,J,l,t] \right),
\end{split}
\end{equation}
 where $C$ is the following functional of the density:
 \begin{equation}
 C[\rho, J,l,t] =  \int_0^t \frac{-l-\beta \rho(s) q(l)}{l(1-J)-  (1+\zeta - l J )e^{ ( l-1-\zeta) (s-t)}} ds.
\end{equation}

\subsection{Closure condition on the density}

 The definition of the density at time $t$, Eq. \ref{densities}, induces the 
  following consistency condition, based on the derivative of the generating function w.r.t. $J$, 
  expressed in the functional form we have just obtained in Eq. \ref{logH}:
\begin{equation}\label{consistency}
\begin{split}
  \rho(t) = &\int_0^1 dl \frac{\partial}{\partial J}|_{J=1}\left[  H\left( 1- \frac{(1-J)(l-1-\zeta)}{ l(1-J)-\left( l(1-J) + (1+\zeta - l) \right) e^{(1+\zeta - l )t}}, l, 0 \right) \right]\\
   &+ \int_0^1 dl  H( 1, l, 0 )  \frac{\partial}{\partial J}|_{J=1}\left( \exp\left((l-1-\zeta)( 1 - J)  C[\rho,J,l,t] \right) \right).
\end{split}
\end{equation}
  The normalisation condition of the probability distribution $p_l$  of occupation numbers at level $l$ reads (at time $0$):
\begin{equation}
H(1,l,0) = 1.
\end{equation}
  The derivative needed in the integrand of the 
 first term in Eq. \ref{consistency} is obtained from a Taylor expansion at first order in $h$ for $J=1+h$ {{(for negative $h$)}}:
\begin{equation}
\begin{split}
H\left( 1- \frac{-h(l-1-\zeta)}{-hl -\left( -hl+ (1+\zeta - l) \right) e^{(1+\zeta - l )t}}, l, 0 \right) =& H(1,l,0)  \\
   &+e^{-(1+\zeta-l)t} h\left(\frac{\partial}{\partial J}|_{J=1}  H\left( J, l, 0 \right)\right)+o(h) \\
=& \;1+ e^{-(1+\zeta-l)t}h \overline{n_l}(0) + o(h),
\end{split}
\end{equation}
We read off
\begin{equation}
\frac{\partial}{\partial J}|_{J=1}\left[  H\left( 1- \frac{(1-J)(l-1-\zeta)}{ l(1-J)-\left( l(1-J) + (1+\zeta - l) \right) e^{(1+\zeta - l )t}}, l, 0 \right) \right]
 = e^{-(1+\zeta-l)t} \overline{n_l}(0).
\end{equation}
On the other hand, the factor of $(1-J)$ in the argument of the exponential of the second term of Eq. \ref{consistency}
 yields immediately 
\begin{equation}
 \frac{\partial}{\partial J}|_{J=1}\left( \exp\left((l-1-\zeta)( 1 - J)  C[\rho,J,l,t] \right) \right)= -(l-1-\zeta)  C[\rho,1,l,t].
\end{equation}
We therefore find the following  closure condition on the density:
\begin{equation}
\rho(t) =\int_0^1 e^{-(1+\zeta - l )t} \overline{n_l}(0) dl - \int_0^1  (l-1-\zeta)  C[\rho,1,l,t] dl.
\end{equation}
This integral  condition is a Volterra  equation:
 \begin{equation}
  \rho(t) =   z( t )  + ( K\ast \rho)(t),
\end{equation}
where the function $z$ contains the initial conditions
 \begin{equation}\label{zDef}
   z( t )  :=  \int_0^1 e^{-(1+\zeta - l )t} \overline{n_l}(0) dl +  \int_0^1\left( \int_0^t ds  l e^{(l-1 - \zeta)(t-s)} \right) dl,
\end{equation}
and the kernel $K$ depends only on the density $q$  and the rates of the three stochastic processes:
\begin{equation}\label{KDef}
 K(T) := \beta \int_0^1  q(l) e^{(l-1 - \zeta)T} dl.
\end{equation}

Let us denote the Laplace transform of a function $f$ of time as follows:
\begin{equation}
 \widetilde{f}( s ) =  \int_0^\infty   e^{-st} f(t)dt.
\end{equation}
  The Laplace transform of the convolution equation satisfied by the density yields 
 the following expression for the Laplace transform of the density, in terms of the parameters 
 of the process and the initial conditions:
 \begin{equation}\label{LaplaceRho}
  \widetilde{\rho}(s) =   \frac{ \widetilde{z}(s)}{ 1- \widetilde{K}(s)}.
\end{equation}

\subsection{Expectation value of the occupation number at a fixed fitness level}

We can adapt the derivation of the consistency equation (Eq. \ref{consistency}) 
 to obtain the expectation value $\overline{n_l}$ of the number of particles at a fixed fitness level $l$:
\begin{equation}
\begin{split}
  \overline{n_l}(t) = &  \frac{\partial}{\partial J}|_{J=1} H( J, l, t )\\
=& \frac{\partial}{\partial J}|_{J=1}\left[  H\left( 1- \frac{(1-J)(l-1-\zeta)}{ l(1-J)-\left( l(1-J) + (1+\zeta - l) \right) e^{(1+\zeta - l )t}}, l, 0 \right) \right]\\
 & +   H( 1, l, 0 )  \frac{\partial}{\partial J}|_{J=1}\left( \exp\left((l-1-\zeta)( 1 - J)  C[\rho,J,l,t] \right) \right)\\
=&  e^{-(1+\zeta - l )t} \overline{n_l}(0)  - (l-1-\zeta)  C[\rho,1,l,t] \\
=&  e^{-(1+\zeta - l )t} \overline{n_l}(0)  + \left( \int_0^t ds ( l+\beta \rho(s) q(l)) e^{(l-1 - \zeta)(t-s)} \right).\\
\end{split}
\end{equation}
 The expectation value of the occupation number at fitness level $l$ therefore forgets
 the initial condition $\overline{n_l}(0)$ at an exponential rate, which decreases linearly with the fitness:
\begin{equation}
 \overline{n_l}(t) = e^{-(1+\zeta - l )t} \overline{n_l}(0) + \frac{l- l e^{-(1+\zeta - l)t)}}{1+\zeta -l} +  \beta q(l) \int_0^t e^{(l-1 - \zeta)(t-s)}\rho(s) ds.
\end{equation}

 The Laplace transform of exponential functions, 
 \begin{equation}
  \int_0^\infty  e^{-st} e^{at} dt = \frac{1}{s-a},
\end{equation}
 yields the following expression for the Laplace transform of the  average occupation number at fitness level $l$, for $s>0$:
 \begin{equation}\label{meanLaplace}
  \widetilde{ \overline{n}_l}(s) =   \left(  \overline{n_l}(0)  - \frac{l}{{1+\zeta -l}}   \right)\frac{1}{s+1+\zeta - l} +  \frac{l}{1+\zeta -l}\frac{1}{s} + 
     \beta \frac{q(l)}{ s+1+\zeta - l}  \widetilde{\rho}(s).
\end{equation}

 As we have expressed the Laplace transform $\widetilde{\rho}$ of the density
  in Eq. \ref{LaplaceRho}, the expectation value $\overline{n_l}$ is 
 completely characterised by Eq. \ref{meanLaplace} in the Laplace domain. Inverting  the Laplace 
 transform of the density would yield an expression for $\overline{n_l}$ as a function of time. However, the 
  explicit result we have obtained in the Laplace domain  
  has a direct application to a system undergoing resetting at random  times distributed with a
 constant rate.

%
%


\section{Resetting events}\label{resetting}

\subsection{Review of the renewal argument}
  Let us consider the  system of particles evolving according to the processes we described in Section \ref{model},
  but assume it is reset to its initial configuration at random times. This configuration 
  is given by the number of particles in every state, which induces the collection of occupation numbers
  at every fitness level.\\
  
Assume the resetting events occur at a fixed rate $r$. The configurations of the 
 system under resetting are the same as the ones in the system with no resettting, but 
 they have different probabilities. For a fixed configuration, let us denote by $\mathcal{P}_r(t)$
 the probability of this configuration in the system undergoing resetting at rate $r$ (and by $\mathcal{P}_0(t)$
the probability of this configuration in the system without resetting), conditional on the fixed initial configuration 
 to which the system is instantaneously brought back at each resetting event. 
 Following the renewal argument of  \cite{evans2018run,stochasticReview}, the probability distribution $\mathcal{P}_r(t)$ 
 can be expressed in terms of the probability distribution  $\mathcal{P}_0$.
  The expression consists of two terms corresponding to the following alternative. 
  There has  been either no resetting event in the time interval $[0,t]$ (the probability 
 of this event is $e^{-rt}$), or the last resetting event occurred within an infinitesimal
 time $d\tau$ of  $t - \tau$, for some 
 value of $\tau$ in $[0,t]$ (the probability density of this event is $re^{-rt}dt$).
  Conditioning on these events yields:
\begin{equation}\label{renewalRev}
   \mathcal{P}_r(t)= e^{-rt} \mathcal{P}_0(t)+ r \int_0^t e^{-r\tau} \mathcal{P}_0(\tau) d\tau.
\end{equation}
  The large-time limit of this equation shows that the Laplace transform of  the ordinary probability distribution 
  yields the steady-state  probability distribution of the system under resetting.

\subsection{Application to the steady state of the system under resetting}
 Let us apply this very general argument to our model. 
 Consider a fitness level $l$ and an integer $k\geq 0 $, and denote by $P_0(k,t,l)$ the probability 
 of the presence of $k$ particles at fitness level $l$ at time $t$ (conditional on a fixed
  configuration at time zero), in the absence of resetting. 
 Let us denote by $P_r(k,t,l)$ the probability of the same event (conditional on the same initial condition),
 in the presence of resetting at rate $r$. With this notation
 the probability distribution satisfying the master equation (Eq. \ref{evolPDE})
  reads $p_l(k,t) = P_0(k,t,l)$.\\

 Rewriting the renewal equation \ref{renewalRev} in our notations,
 we obtain:  
 \begin{equation}
  P_r( k,t,l) = e^{-rt} P_0(k,t,l) + r \int_0^t e^{-r\tau}P_0(k,\tau,l) d\tau,\;\;\;\;\forall k\geq 0,\;\;\;\forall l \in (0,1).
 \end{equation}
The expectation value of the occupation number of  fitness level $l$ under resetting is  
 therefore expressed  as:
 \begin{equation}\label{meanReset}
 \sum_{k\geq 0} k P_r( k,t,l) = e^{-rt} \sum_{k\geq 0} k P_0(k,t,l) + r \int_0^t e^{-r\tau}\sum_{k\geq 0} P_0(k,\tau,l) d\tau.
 \end{equation}
 Let us   introduce the following notation: 
 \begin{equation}
  \overline{n}_{l,r}(t) :=  \sum_{k\geq 0} k P_r( k,t,l).
 \end{equation}
  Let us  keep the notation $\overline{n}_{l}$ introduced in Eq. \ref{densities}
 for the expectation value of the occupation number at level $l$ 
 in the absence of resetting (even though the notation  $\overline{n}_{l,0}$ would also be consistent 
 for this quantity), and rewrite the renewal equation (Eq. \ref{meanReset}):
\begin{equation}\label{meanReset}
 \overline{n}_{l,r}(t)= e^{-rt} \overline{n}_{l}(t) + r \int_0^t e^{-r\tau} \overline{n}_{l}(\tau) d\tau.
 \end{equation}

 Consider the steady state of the system under resetting. The expectation value of the number of particles at fitness level $l$ 
 is read off from the large-time limit of Eq. \ref{meanReset}
 in terms  the Laplace transform of $\overline{n}_{l}$:
\begin{equation}\label{steadyDensity}
  \overline{n}_{l,r}(\infty) = r \widetilde{\overline{n}_{l,r}}(r).
 \end{equation}

 The expression needed on the r.h.s. was worked out in Eq. \ref{meanLaplace} when we took the Laplace 
 transform of the convolution equation satisfied 
 by $\overline{n}_{l}$. We therefore read off
\begin{equation}
  \overline{n}_{l,r}(\infty) = \left(  \overline{n_l}(0)  - \frac{l}{{1+\zeta -l}}   \right)\frac{r}{r+1+\zeta - l} +  \frac{l}{1+\zeta -l} + 
     \beta \frac{q(l)}{ r+1+\zeta - l}  r\widetilde{\rho}(r).
 \end{equation}
Integrating Eq. \ref{steadyDensity} w.r.t.  fitness  yields the expression of the steady-state density 
 $\rho_r(\infty)$ in terms of the Laplace transform of the ordinary density:
\begin{equation}\label{steadyDensity}
  \rho_r(\infty) = \int_0^1 \overline{n}_{l,r}(\infty) dl = r\int_0^1 \widetilde{\overline{n}_{l}}(r) dl = r\widetilde{\rho}(r).
 \end{equation}
The average occupation number  at level $l$ in the steady state of the system under resetting contains a skewed version 
 of the density $q$, weighted by the hopping rate $\beta$ and  the average steady-state density of the system. Moreover,
 the resetting rate combines with the sum of hopping and death rates, to yield a factor with the same structure as in the
  steady state without resetting (Eq. \ref{steadyOrdinary}):
\begin{equation}\label{steadyReset}
  \overline{n}_{l,r}(\infty) =  \frac{l + r \overline{n_l}(0)}{1+\zeta +r -l} + 
     \frac{ \beta\rho_r(\infty)q(l)}{ 1+\zeta +r - l}.
 \end{equation}
 
 The  density parameter $\rho_r(\infty)$ can be expressed using the consistency condition,
 \begin{equation}
 \rho_r(\infty) = \int_0^1 \overline{n}_{l,r}(\infty) dl,
 \end{equation}
 which yields 
\begin{equation}\label{steadyReset}
  \rho_r(\infty)  =    \int_0^1 \frac{l +  r \overline{n_l}(0)}{1+\zeta +r -l} dl\left( 1 - \beta \int_0^1 \frac{q(l) dl}{1+\zeta + r -l}\right)^{-1}.  
 \end{equation}
 If $\beta$ is below the critical value  worked out in the absence of resetting (Eq. \ref{critical}),
 the denominator in the above quantity is  positive. 
 On the other hand, the density $\rho_r(\infty)$ can be predicted by 
 the Laplace transform of the Volterra equation, using Eqs \ref{zDef},\ref{KDef},\ref{LaplaceRho}:
\begin{equation}
\widetilde{z}(r) = \int_0^1 \frac{\overline{n_l}(0)dl }{1+\zeta + r -l} + \int_0^1 \left(\frac{1}{r}- \frac{1}{r+1+\zeta -l} \right)\frac{ldl}{1+\zeta  -l}=\int_0^1 \frac{\overline{n_l}(0)dl }{1+\zeta + r -l} + \int_0^1 \frac{1}{r}\frac{ldl}{1+\zeta + r -l},
\end{equation}
\begin{equation}
\widetilde{K}(r) = \beta \int_0^1 \frac{q(l)dl}{1+\zeta+ r - l},
\end{equation}
\begin{equation}
r \widetilde{\rho}(r) = \frac{1}{1-\beta  \int_0^1 \frac{q(l)dl}{1+\zeta+ r - l}}
 \int_0^1 \frac{(r \overline{n_l}(0) +l)dl}{1+\zeta + r -l},
\end{equation}
 which  confirms the value obtained from the consistency condition (using Eq. \ref{steadyDensity} relating the steady-state density 
 to the Laplace transform of $\rho$ at the resetting rate).

\section{Conclusions}

We have expressed the generating function of a non-conserving ZRP with extensive rates
  as a functional of the average density.  The density  satisfies an integral equation 
 of Volterra type.  This equation implies that the expectation value of the 
  occupation number of a given fitness level goes to its steady-state value at an exponential rate that decreases linearly with the 
 fitness. Moreover, the Volterra equation satisfied by the density  
  is enough to characterise the expectation value of the occupation number at every fitness level
 in the Laplace domain.\\


   We obtained the 
 expectation values of the occupation numbers at each fitness level of the 
 system under stochastic resetting at {{constant}} rate. They are induced by the values of the Laplace transform of the 
  ordinary occupation numbers, evaluated at the resetting rate $r$. The result has a very simple structure 
  that interpolates between the initial conditions (at strong resetting) and the ordinary case (at zero resetting).
 {{The hopping process  can be mapped to a set of independent random walkers on a fully connected, 
 large set of states. Recent developments on stochastic resetting
of interacting particle systems have addressed properties of the symmetric exclusion
 process \cite{basu2019symmetric} and totally asymmetric exclusion 
  process  \cite{karthika2020totally}, and in \cite{magoni2020ising}, the phase diagram in the plane of temperature 
 and resetting rate has been presented for the Ising model,
an interacting system with a thermodynamic  phase transition
 in its equilibrium state. However, the present model allows for
 a mapping of the model of growing networks, as well 
 as to a microscopic version of Kingman's house-of-cards model.}}\\

  Indeed the system  can be interpreted as  a model of the balance of selection and mutation, 
     or (by mapping particles to oriented links), as a model of a network in which the links can form at a rate given by the 
 fitness level, vanish at a fixed rate $\delta$, and be randomly  rewired at a rate $\beta$. The model then becomes a version of the Bianconi--Barab\'asi model with random 
 rewiring of the destination of links.
The ordinary case, worked out in \cite{ZRPSS},
  corresponds to a skewed version of the density of sates $q$, with an extra term 
 that gives rise to an atom at maximum density when  the sum of the death and hopping rates
  approaches the maximum production rate (the production rate of particles at maximum fitness,
 which we set to $1$).
  The resetting process modifies the ordinary formula in two intuitive ways, that are enough to completely
 recover the result:\\
1. the resetting rate $r$  adds up to the sum of the death and mutation rates  (the combination denoted by $1+\zeta$ 
 in Eq. \ref{zetaDef}), as one of the processes balancing selection, so that an atom develops at maximum fitness in 
 the limit of zero resetting rate and zero $\zeta$ (in this limit the steady-state density of the system diverges as $-\log(\zeta+r)$);\\
2. the initial conditions add up to the  term $\beta\rho_r(\infty)q(l)$ through the combination $r \overline{n_l}(0)$.\\
 The steady-state density $\rho_r(\infty)$ can be recovered as a consistency condition after applying 
 these two prescriptions to the ordinary expression (even though we obtained it from the Laplace transform of the 
 Volterra equation in the first place). The resetting process at constant 
   rate  therefore modifies the house-of-cards effect in the following sense: it combines with the 
 mutation to reshuffle the genomic deck, mixing the mutant density with the initial conditions.\\

 {{Technically, the Laplace transform of the average occupation numbers was enough to characterise the 
 steady state of the model under resetting, because of the Poisson distribution of the resetting times on which we condition to work out 
  the renewal equation. Inverting the Laplace transform  would allow to study the large-time behaviour of the integral
 of the occupation numbers on a shrinking interval of high fitnesses of the form $[1- x/t, 1]$. This behaviour would depend on the effective annihilation rate 
 $\beta+\delta$. It would be interesting to let this parameter go to $1$ (which leads to the emergence of an atom at mximum fitness due to the first integral in Eq. \ref{steadyOrdinary}), and to see whether the large-time limit of the total population in the considered shrinking interval could be expressed 
 as a function of $x$ only. This is the case in the Kingman house-of-cards model, in which this function is related to the incomplete
 gamma function, provided the mutant density vanishes as a power law at maximum fitness \cite{dereich2013emergence}. This gamma-wave
 shape has been shown to occur in other  processes exhibiting condensation \cite{dereich2016preferential,mailler2016condensation}.}}\\

{{The  resetting events  we considered leave the set of states fixed. In population dynamics, this corresponds to a fixed mutant density. 
 We assumed this density to vanish at maximum fitness, which is equivalent to a 
 vanishing probability of a beneficial mutation at high values of fitness. The condensate in the house-of-cards model \cite{KingmanSimple} is  indeed
 an effect of selection only. 
 However, the emergence of beneficial mutations can be modelled by new values of fitness drawn from a mutant density  with an unbounded support.
 Such a model was studied in  \cite{park2008evolution} for a measure-valued model in the limit of an  infinite population. Adapting the present model 
 to such a mutant density would require to give a prescription for the gradual introduction of states in the system, because starting with states
  of arbitrarily high reproductive fitness would quickly  give rise to divergences in the occupation numbers. Moreover, a more radical resetting prescription would involve 
 a complete destruction of particles and states. The hopping process would take place on a growing lattice that would approach the underlying 
  mutant density, until the next resetting event interrupts this growth process. }}
 



\section*{Acknowledgements}

It is a pleasure to thank Linglong Yuan for numerous discussions.

\bibliography{bibRefsNewRefixed} 
\bibliographystyle{ieeetr}
\end{document}